\documentclass[twocolumn]{aastex63}

\newcommand{\msun}{\ifmmode M_{\odot}\else$M_{\odot}$\fi}
\newcommand{\rsun}{\ifmmode R_{\odot}\else$R_{\odot}$\fi}
\newcommand{\degrees}{\ifmmode^{\circ}\else$^{\circ}$\fi}
\newcommand{\amin}{\ifmmode^{\prime}\else$^{\prime}$\fi}
\newcommand{\asec}{\ifmmode^{\prime\prime}\else$^{\prime\prime}$\fi}

\received{November 10, 2021}
\revised{December 28, 2021}
\accepted{December 28, 2021}

\submitjournal{Astrophysical Journal}

\shorttitle{Two New Pulsars in M28}
\graphicspath{{./}{figures/}}
\usepackage{hyperref}
\begin{document}
\shortauthors{Douglas et al.}

\title{Two New Black Widow Millisecond Pulsars In M28}

\email{adouglas1298@gmail.com}

\author[0000-0001-5492-4215]{Andrew Douglas}
\affiliation{University of Virginia, 1826 University Avenue, Charlottesville, VA 22904, USA}
\author[0000-0001-5624-4635]{Prajwal V.~Padmanabh}
\affiliation{Max-Planck-Institut f\"{u}r Radioastronomie, Auf dem H\"{u}gel 69, D-53121 Bonn, Germany}
\author[0000-0001-5799-9714]{Scott M.~Ransom}
\affiliation{NRAO, 520 Edgemont Road., Charlottesville, VA 22903, USA}
\author[0000-0001-6762-2638]{Alessandro Ridolfi}
\affiliation{ INAF -- Osservatorio Astronomico di Cagliari, Via della Scienza 5, I-09047 Selargius (CA), Italy}
\affiliation{Max-Planck-Institut f\"{u}r Radioastronomie, Auf dem H\"{u}gel 69, D-53121 Bonn, Germany}
\author[0000-0003-1307-9435]{Paulo Freire}
\affiliation{Max-Planck-Institut f\"{u}r Radioastronomie, Auf dem H\"{u}gel 69, D-53121 Bonn, Germany}
\author[0000-0001-9518-9819]{Vivek Venkatraman Krishnan}
\affiliation{Max-Planck-Institut f\"{u}r Radioastronomie, Auf dem H\"{u}gel 69, D-53121 Bonn, Germany}
\author[0000-0001-8715-9628]{Ewan~D.~Barr}
\affiliation{Max-Planck-Institut f\"{u}r Radioastronomie, Auf dem H\"{u}gel 69, D-53121 Bonn, Germany}
\author[0000-0002-7104-2107]{Cristina Pallanca}
\affil{Dipartimento di Fisica e Astronomia “Augusto Righi'', Università degli Studi di Bologna, Via Gobetti 93/2, 40129 Bologna, Italy}
\affil{INAF-Osservatorio di Astrofisica e Scienze dello Spazio di Bologna, Via Gobetti 93/3 I-40129 Bologna, Italy}
\author[0000-0002-5038-3914]{Mario Cadelano}
\affil{Dipartimento di Fisica e Astronomia “Augusto Righi'', Università degli Studi di Bologna, Via Gobetti 93/2, 40129 Bologna, Italy}
\affil{INAF-Osservatorio di Astrofisica e Scienze dello Spazio di Bologna, Via Gobetti 93/3 I-40129 Bologna, Italy}
\author[0000-0001-5902-3731]{Andrea Possenti}
\affiliation{ INAF -- Osservatorio Astronomico di Cagliari, Via della Scienza 5, I-09047 Selargius (CA), Italy}
\author[0000-0001-9784-8670]{Ingrid Stairs}
\affiliation{Dept. of Physics and Astronomy, University of British Columbia, 6224 Agricultural Road, Vancouver, BC V6T 1Z1, Canada}
\author[0000-0003-2317-1446]{Jason W.~T.~Hessels}
\affiliation{ASTRON, the Netherlands Institute for Radio Astronomy, Oude Hoogeveensedijk 4, 7991 PD Dwingeloo, The Netherlands}
\affiliation{Anton Pannekoek Institute for Astronomy, University of Amsterdam, Postbus 94249, 1090 GE Amsterdam, The Netherlands}
\author[0000-0002-2185-1790]{Megan E. DeCesar}
\affiliation{George Mason University, Fairfax, VA 22030, resident at the U.S. Naval Research Laboratory, Washington, D.C. 20375, USA}
\author[0000-0001-5229-7430]{Ryan S.\ Lynch}
\affiliation{Green Bank Observatory, PO Box 2, Green Bank, West Virginia, 24944, USA}
\author[0000-0003-3294-3081]{Matthew Bailes}
\affiliation{Centre for Astrophysics and Supercomputing/OzGrav, Swinburne University of Technology, Mail 74, PO Box 218, Hawthorn,
Vic, 3122, Australia}
\author[0000-0002-8265-4344]{Marta Burgay}
\affiliation{ INAF -- Osservatorio Astronomico di Cagliari, Via della Scienza 5, I-09047 Selargius (CA), Italy}
\author[0000-0003-1361-7723]{David~J.~Champion}
\affiliation{Max-Planck-Institut f\"{u}r Radioastronomie, Auf dem H\"{u}gel 69, 53121 Bonn, Germany}
\author[0000-0002-5307-2919]{Ramesh Karuppusamy}
\affiliation{Max-Planck-Institut f\"{u}r Radioastronomie, Auf dem H\"{u}gel 69, D-53121 Bonn, Germany}
\author[0000-0002-4175-2271]{Michael Kramer}
\affiliation{Max-Planck-Institut f\"{u}r Radioastronomie, Auf dem H\"{u}gel 69, D-53121 Bonn, Germany}
\author[0000-0001-9242-7041]{Benjamin Stappers}
\affiliation{Jodrell Bank Centre for Astrophysics, Department of Physics and Astronomy, The University of Manchester, Manchester M13 9PL, UK}
\author[0000-0003-0292-4453]{Laila Vleeschower}
\affiliation{Jodrell Bank Centre for Astrophysics, Department of Physics and Astronomy, The University of Manchester, Manchester M13 9PL, UK}



\begin{abstract}

We report the discovery of two Black Widow millisecond pulsars in the globular cluster M28 with the MeerKAT telescope. PSR J1824$-$2452M (M28M) is a 4.78-ms pulsar in a 5.82\,hour orbit and PSR J1824$-$2452N (M28N) is a 3.35-ms pulsar in a 4.76\,hour orbit. Both pulsars have dispersion measures near 119.30\,pc\,cm\textsuperscript{-3} and have low mass companion stars ($\sim$0.01-0.03\,\msun), which do not cause strong radio eclipses or orbital variations. Including these systems, there are now five known black widow pulsars in M28. The pulsar searches were conducted as a part of an initial phase of MeerKAT's globular cluster census (within the TRAPUM Large Survey Project). These faint discoveries demonstrate the advantages of MeerKAT's survey sensitivity over previous searches and we expect to find additional pulsars in continued searches of this cluster.
\end{abstract}

\keywords{M28 -- Pulsar -- PRESTO -- Jerk Search -- MeerKAT -- MeerTime -- TRAPUM}

\section{Introduction} \label{sec:intro}

Globular clusters (GCs) are rich environments for the creation of millisecond pulsars \citep[MSPs; e.g.][]{Ransom2008}. MSPs are created when neutron stars
accrete mass and angular momentum from companion stars and are spun-up to millisecond periods.
Due to the high stellar density in the central regions of globular clusters between 1,000 and 100,000\,\msun\,pc$^{-3}$ \footnote{Structural Parameters of Globular Clusters: \url{https://people.smp.uq.edu.au/HolgerBaumgardt/globular/parameter.html}} stellar interactions produce dynamically-formed binaries that can spin-up neutron stars \citep[e.g.~][]{Verbunt2014}. Additionally, these dynamical encounters can produce MSPs in highly eccentric and exotic orbits. Due to the high likelihood of finding MSPs within GCs, and especially those in exotic orbits, GCs have been the focus of many pulsar searches using the largest radio telescopes and over 232 have been found so far\footnote{List of pulsars in globular clusters: \url{http://www.naic.edu/~pfreire/GCpsr.html}}. There are 39 pulsars currently known in Terzan~5 \citep{lyne1990_ter5,lyne2000_ter5,ransom2005,Hessels2006_ter5,andersen2018,cadelano2018_ter5,Ridolfi2021} and 27 in 47~Tucanae \citep{Camilo2000,Manchester90_47tuc,Robinson1995_47tuc,Pan2016_47tuc,Ridolfi2021}. M28 had, up until this work, 12 known pulsars \citep{beg06}, ranking it as the third cluster in number of known pulsars.\footnote{Note that NGC~1851 is now tied with M28: \url{http://www.naic.edu/~pfreire/GCpsr.html}}

M28 (NGC~6626) is a fairly massive globular cluster in the bulge of the Milky Way at a distance of $\sim$5.4\,kpc \citep{bv2021}. It contains the first known globular cluster pulsar \citep[PSR B1821$-$24 or M28A;][]{lyne1987}, as well as a transitional MSP that behaves alternatively as a radio MSP and as an active X-ray binary \citep[IGR J18245$–$2452 or M28I;][]{Papitto2013}. M28 has been the subject of many multi-wavelength searches in the past few decades. {\em ROSAT} observed M28A as a possible x-ray source \citep{Danner1994} and the {\em Chandra X-ray Observatory} has observed the cluster multiple times and in 2011 confidently located Pulsars A, C, D, F, H, J and K in X-rays. \citep[e.g.~][]{Bogdanov2011}. The Green Bank Telescope (GBT) also observed M28 multiple times and uncovered 11 radio MSPs in the early to mid 2000s \citep{beg06}. One notable point about these MSPs is that they seem to contain a disproportionately large percentage of Black Widow pulsars, i.e. systems where the companion star is of very low mass ($\leq$ 0.04\,\msun), often causing eclipses of the radio pulsations, and having orbits of 2$-$10\,hrs in duration \citep{Roberts2013}. M28 also contains two 
Redback pulsars, i.e. pulsar binaries with a more massive companion ($\sim$ 0.1$-$0.2\,\msun) than the Black Widow systems, but also showing eclipses or timing irregularities, sometime across the entire orbit: Black Widow and Redback pulsars are collectively dubbed as {\it spiders} \citep{Roberts2013}.  

MeerKAT is a newly commissioned 64-dish radio telescope array in South Africa \citep{Booth2012}. Two pulsar-based large science projects, MeerTime and TRAPUM, working together as described below, have made observations of several globular clusters. Initial results from a globular cluster survey targeting clusters with known pulsars, for the purposes of comparing older detections with MeerKAT ones, has been presented by \citet{Ridolfi2021}. With 15 years since the last dedicated searches (using the GBT), the high sensitivity of MeerKAT coupled with better computing power and the availability of higher-dimension acceleration or ``jerk'' searches \citep[e.g.~][]{andersen2018} motivated new dedicated searches of M28.

In this paper we present the discovery and timing of M28M and M28N. In \S\ref{sec:obsndata} we present the radio observations and detection of the MSPs and the subsequent timing procedures. In \S\ref{sec:optical} we discuss searches for the systems in archival {\em HST} and {\em Chandra} data. In \S\ref{sec:discussion} we discuss some implications of these new systems and in \S\ref{sec:conclusion} we present our conclusions. 
 
\section{Observations and Data Analysis}
\label{sec:obsndata}
\subsection{Data}
\label{sec1.1}
The initial MeerKAT data that we used to search for new pulsars in M28 was part of the series of survey observations using the core MeerKAT antennas ($\sim$42 out of 64 dishes) that targeted globular clusters containing known pulsars \citep{Ridolfi2021}. These observations were made as part of a joint effort between the MeerTime project \citep{bailes2020}, which conducts long-term pulsar timing, and the TRAPUM project \citep[TRansients And PUlsars with MeerKAT;][]{stappers2016}\footnote{TRAPUM: \url{http://trapum.org}}, whose goal is to discover new radio pulsars in targeted regions that have yielded significant discoveries in the past as well as environments that are conducive for pulsar formation.

Two 9000-second duration MeerTime observations occurred on 2019 July 19 and 2020 February 4, with 642\,MHz of nominal bandwidth centered at 1283.58\,MHz and split into 768 frequency channels. Each channel was coherently dedispersed at a dispersion measure (DM) of 119.892\,pc\,cm$^{-3}$, which is roughly the average value of the DMs of the known pulsars in the cluster, and were subsequently further frequency subbanded, using incoherent dedispersion, into 256 total-intensity channels with a sampling time of $\sim$76.56\,$\mu$s. In addition to the two MeerTime observations, we had access to the central beam of two 9000-second TRAPUM observations taken on 2020 December 6 and 11, which were not coherently dedispersed but had 2048 frequency channels over 856\,MHz of bandwidth centered at 1283.86\,MHz. These observations used additional antennas than the earlier MeerTime observations, and were therefore approximately 10$-$20\% more sensitive per unit time than the earlier observations. As with the MeerTime data, these observations were later subbanded, after intra-channel dispersion removal, into 256 frequency channels and integrated in time to $\sim$76.56\,$\mu$s samples for analysis. Analysis of the non-central TRAPUM beams will be published elsewhere.

\subsection{Searches}
\label{sec1.2}
We used {\tt PRESTO} \citep{ransom2011} to perform Fourier domain acceleration and jerk searches on the M28 data using its routine {\tt accelsearch} \citep{2001PhDT.......123R}. The jerk search functionality is a relatively recent addition to {\tt PRESTO} \citep{andersen2018}. Jerk searching aims to find pulsars whose accelerations are changing linearly during an observation. The jerk term improves search sensitivities to binary pulsars when the integration times (or duration of searched segments) are between 5$-$15\% of the orbital periods of the pulsars \citep{Bagchi_2013,andersen2018}.

We searched a range of dispersion measures (DMs) from 115.0\,pc\,cm$^{-3}$ to 125.0\,pc\,cm$^{-3}$, with a step size of 0.1\,pc\,cm$^{-3}$. We chose this range as it is roughly centered on the average DM of the pulsars in the cluster, and it would be unlikely that we would find any MSPs significantly outside this range. We conducted differing levels of maximal acceleration and jerk searching (i.e.~up to {\tt-zmax 200 -wmax 600}), where $z$ represents the number of Fourier bins a signal can drifted during an observation and $w$ is the derivative of that. We also conducted acceleration-only searches on shorter segments of the MeerTime data with 29 and 58 million samples, corresponding to around one third (50 minutes) and two thirds (100 minutes) of the 117 million sample data file, respectively. This segmented method of searching allows one to detect more compact orbits that would be missed with full-duration acceleration searches. With the TRAPUM data, we conducted segmented searches of 45 and 90 million samples, very roughly one quarter (60 minutes) and one half (120 minutes) of the 187 million sample data file, respectively.

We first searched the full-duration observations and segments with acceleration only ({\tt-zmax 200}), hoping to find a pulsar through the increased sensitivity of MeerKAT, but without success. We then switched to jerk searching, with {\tt-zmax 200} and {\tt -wmax 600}. In the full search we detected the new M28M at a DM of 119.30\,pc\,cm$^{-3}$ in the 2019 July 19 MeerTime observation at $z$=19.50 and $w$=10. We were unable to confirm the pulsar in the 2020 February 4 MeerTime observation (until much later, via folding with an orbital ephemeris) and thus searched the more sensitive TRAPUM observations from 2020 December 6th and 11th, and confirmed M28M in those data. We continued searching TRAPUM and MeerTime data at {\tt-zmax 200} and {\tt -wmax 600}, and soon detected a second candidate that we confirmed to be M28N at $z$=3 and $w$=60 in the 2020 February 4 data. Since both pulsars were detected in the central TRAPUM beam (0.246\amin\ radius for the December 6 observation and 0.210\amin\ radius for the December 11 observation), we knew the MSPs must be very close to the cluster center.

\subsection{Timing}
\label{sec1.3}
With multiple detections of both new pulsars over the four MeerKAT observations, we were able to determine circular orbits and average barycentric spin periods for each pulsar using the {\tt fit\_circular\_orbit.py} routine from {\tt PRESTO}. These initial ephemerides allowed us to iteratively fold, detect, and determine pulse arrival times from the large number of archival observations of M28 from the GBT, starting in 2005 and continuing to the present day. We note that the spin period of M28M might actually be 9.56\,ms, twice the period that we report. The pulsar was initially discovered at that period, but folding of the discovery, subsequent, and archival observations at the longer period show two gaussian-like peaks with similar amplitudes and widths, and separated by $\sim$180\degrees.  We therefore timed the pulsar assuming it was a 4.78\,ms rotator. This ambiguity will be resolved  by future MeerTime observations taken with higher time resolution and polarization information.

The GBT data from 2005 to 2008 was a mixture of L-band (centered at 1.5\,GHz) and S-band (centered at 2\,GHz) observations initially using the Pulsar Spigot \citep{kaplan2005} and having effectively $\sim$600\,MHz of available bandwidth with 1536 frequency channels sampled every 81.92\,$\mu$s \citep[see][for more details]{beg06}.  Starting in 2010 and continuing through mid-2020, observations were made with the Green Bank Ultimate Pulsar Processing Instrument \citep[GUPPI;][]{duplain2008}, using the same bands, but with 512 coherently dedispersed channels, subsequently partially integrated to 128 channels with 40.96\,$\mu$s sampling. Finally, for the past year, we used VEGAS \citep{srikanth2012}, in an identical mode to the GUPPI observations.

Both the new pulsars were detected, albeit often with very low signal-to-noise, in the vast majority of the archival GBT observations.  
The total GBT integration time, sampled roughly quarterly, is 49.6\,hours at L-band and 73.8\,hours at S-band.

\begin{figure}[ht!]
\epsscale{1.1}
\plottwo{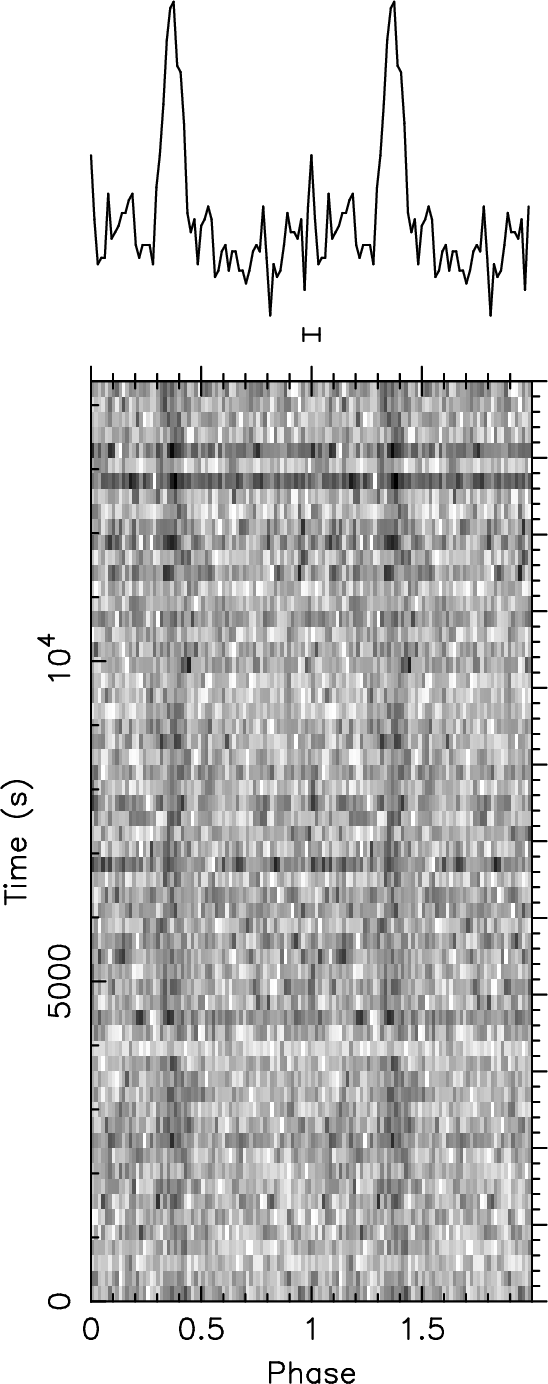}{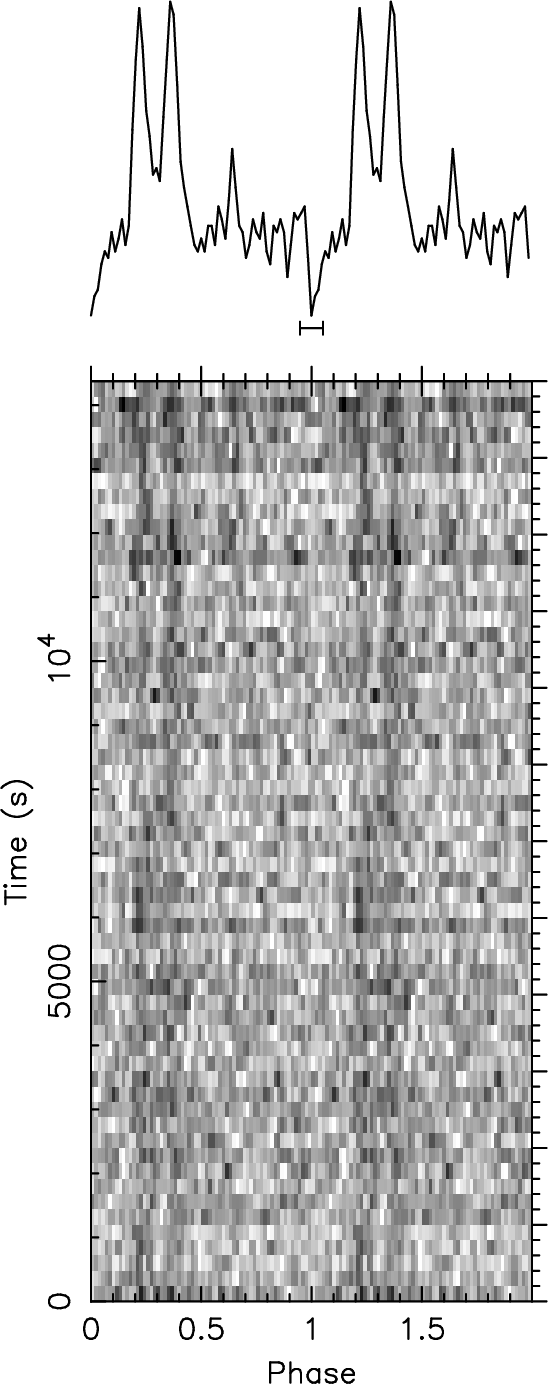}
\caption{The best detections of M28M and M28N, respectively (left and right), from the MeerKAT TRAPUM observation on 2020 December 11. The pulses, each with 64 bins in the profiles, show two full rotations of the pulsar and the greyscale shows the intensity of the emission as a function of time.  The error bar under the summed profiles indicates the approximate time resolution of the data, including dispersive effects, $\sim$350\,$\mu$s. \label{fig:profiles}}
\end{figure}

\begin{figure}[ht!]
\epsscale{1.15}
\plotone{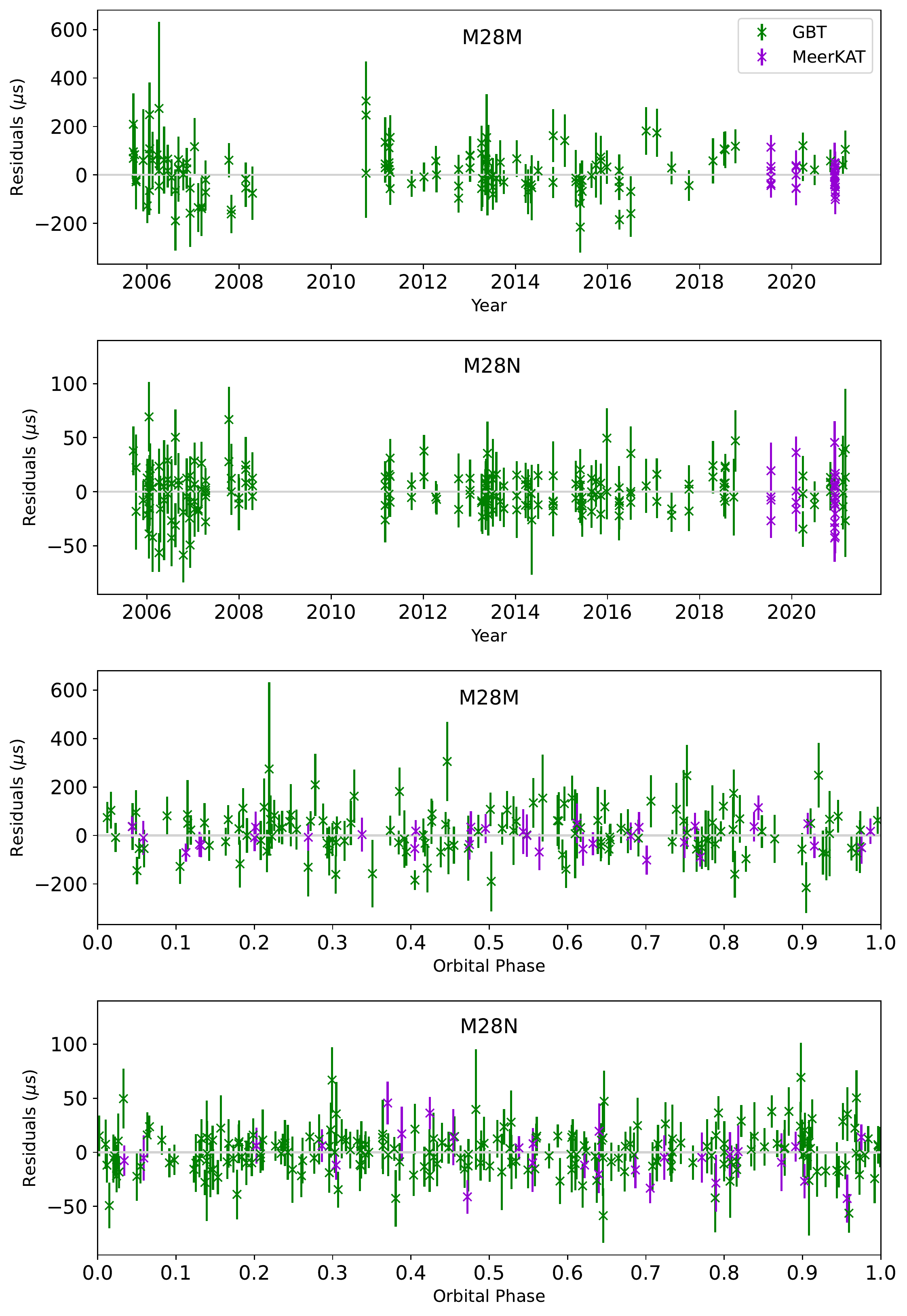}
\caption{Timing residuals for M28M and M28N. The top two panels show timing residuals as a function of time. The bottom two panels show the timing residuals as a function of orbital phase. Eclipses of the pulsed radio emission would typically occur at superior conjunction, which happens at orbital phase 0.25.\label{fig:residuals}}
\end{figure}

We determined times of arrival (TOAs) for each pulsar using the {\tt get\_TOAs.py} routine in {\tt PRESTO} after fitting gaussians with {\tt PRESTO}'s {\tt pygaussfit.py} to the highest signal-to-noise TRAPUM observation (see Fig.~\ref{fig:profiles}) to use as noiseless pulse templates. We were typically able to get roughly one TOA per hour of telescope time with the GBT and one per half hour with MeerKAT. These TOAs enabled us to phase connect each pulsar back to 2005 using {\tt TEMPO} \footnote{\url{http://tempo.sourceforge.net/}}, although the slightly fainter M28M required us to use the {\tt DRACULA}\footnote{\url{https://github.com/pfreire163/Dracula}} software \citep{dracula2018} to establish phase connection. The reduced $\chi^2$ values of the fits for both pulsars are near one (1.20 for M28M and 1.01 for M28N), although for M28M, we needed to increase the nominal TOA errors (i.e.~using the EFAC parameter) by 50\% for Spigot data and 30\% for GUPPI/VEGAS/MeerKAT data to achieve that value.
The final timing solutions, as well as the plots of the associated timing residuals, are provided in Table~\ref{table:timing} and Fig.~\ref{fig:residuals}.

\section{Optical and X-ray observations}
\label{sec:optical}
\subsection{Data set and data reduction}

To search for the optical counterparts of the two newly discovered binaries M28M and M28N, we use the same archival observations used to discover the optical counterparts to M28H and M28I \citep{pallanca10,pallanca13}. The data set is composed of deep and high-resolution optical images acquired with the Wide Field Camera 3 (WFC3) and the Advanced Camera for Surveys (ACS) mounted on the Hubble Space Telescope (HST). We refer to the aforementioned papers for more details about the adopted data set.
The data reduction is performed using DAOPHOT \citep{stetson87} following the prescriptions by \citet{cadelano19,cadelano20b}. 
Instrumental magnitudes are calibrated to the VEGAMAG photometric system using the stars in common with the catalogs presented in \citet{pallanca10,pallanca13}. 
 The instrumental positions, after correction for geometric distortions, are transformed in the absolute coordinate systems using the stars in common with the Gaia Early Data Release 3 \citep{gaia_dr3}. The resulting astrometric uncertainty is about $0.01\arcsec$. We verify the presence of a possible offset between the radio and optical frames using M28H as reference, whose position is very well measured in both frames. The position of M28H in the Gaia optical frame is RA=$18^{h}24^{m}31.61112^{s}$ and Dec=$-24^{\circ}52\arcmin17.39856\arcsec$ and thus differs of about $-0.01\arcsec$ and $0.18\arcsec$ in RA and Dec, respectively, from the radio position quoted in \citet{pallanca10}.  We therefore apply this offset to the optical frame. We cannot include M28I in the offset calculation due to its large uncertainty in the radio timing position.

\subsection{Searching for optical counterparts}

\begin{figure} \centering
\includegraphics[scale=0.22]{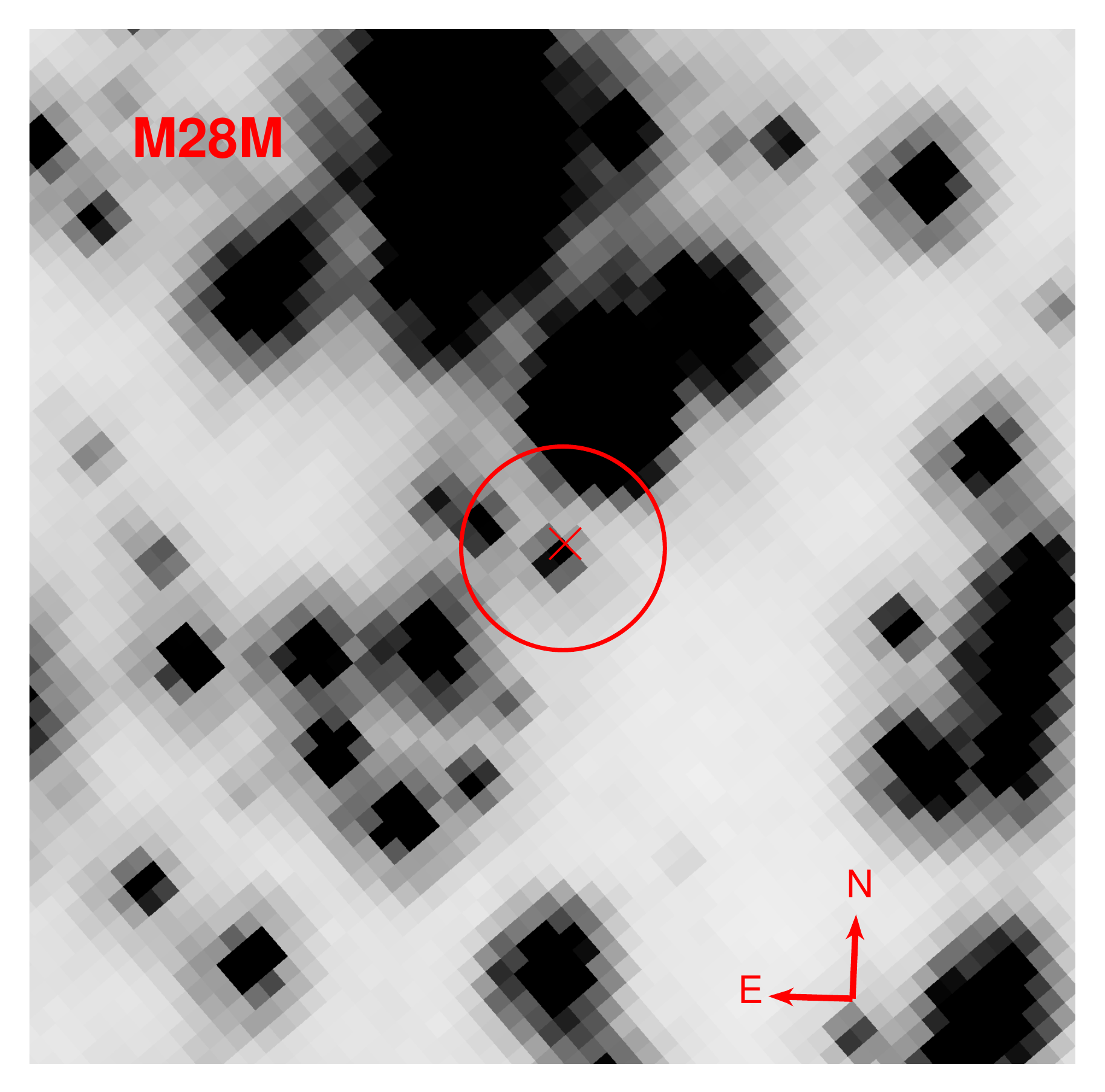}
\includegraphics[scale=0.22]{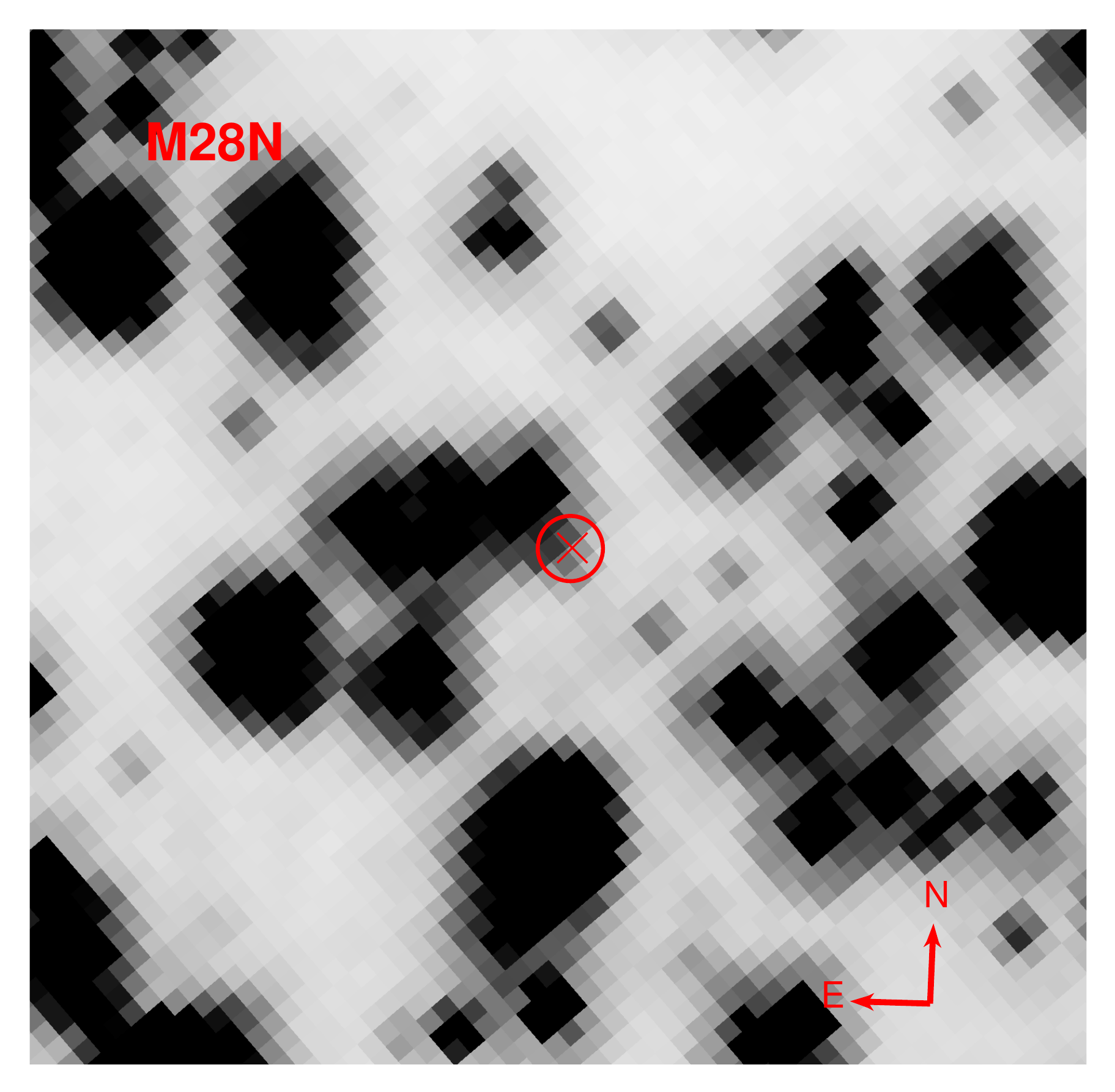}

\caption{{\it Left-hand panel:} $2\arcsec \times 2 \arcsec$ finding chart of the region around the radio position of M28M. The image is a combination of the F814W images. The red cross is centered on the MSP position, and the red circle has a radius equal to the combined radio and optical positional uncertainties. {\it Right-hand panel:} same as in the left-hand panel, but for the case of M28N.}
\label{fig:charts}
\end{figure}

We investigate all of the sources in the area surrounding the positions of the two binaries. In both cases, we find a stellar object in a position compatible, within the positional uncertainty, with that derived from radio data. The finding charts are presented in Figure~\ref{fig:charts}. The candidate optical counterparts to M28M and M28N are displaced from the radio pulsar positions by only $\sim0.05\arcsec$ and $\sim0.07\arcsec$, respectively. 

\begin{figure} \centering
\includegraphics[scale=0.27]{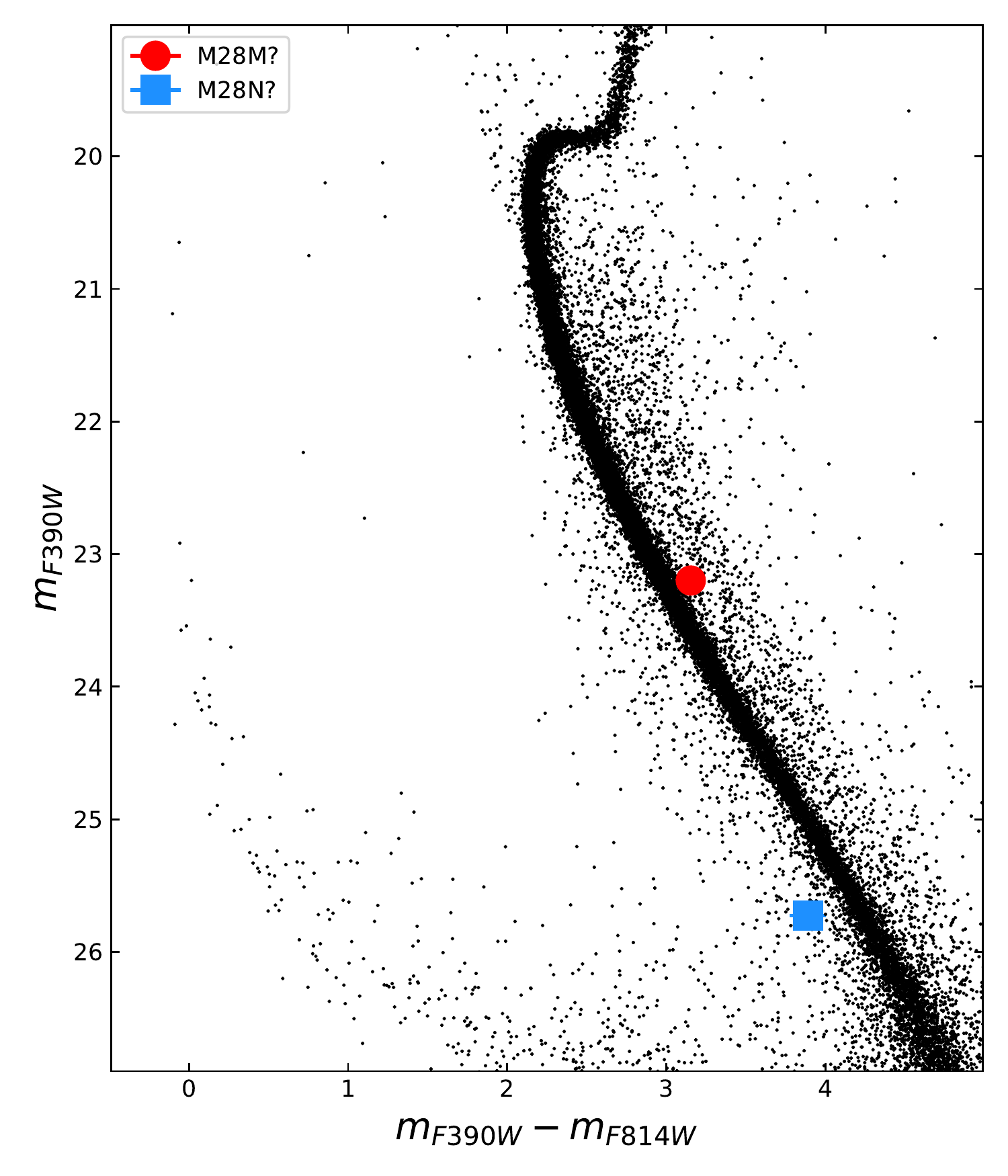}
\includegraphics[scale=0.27]{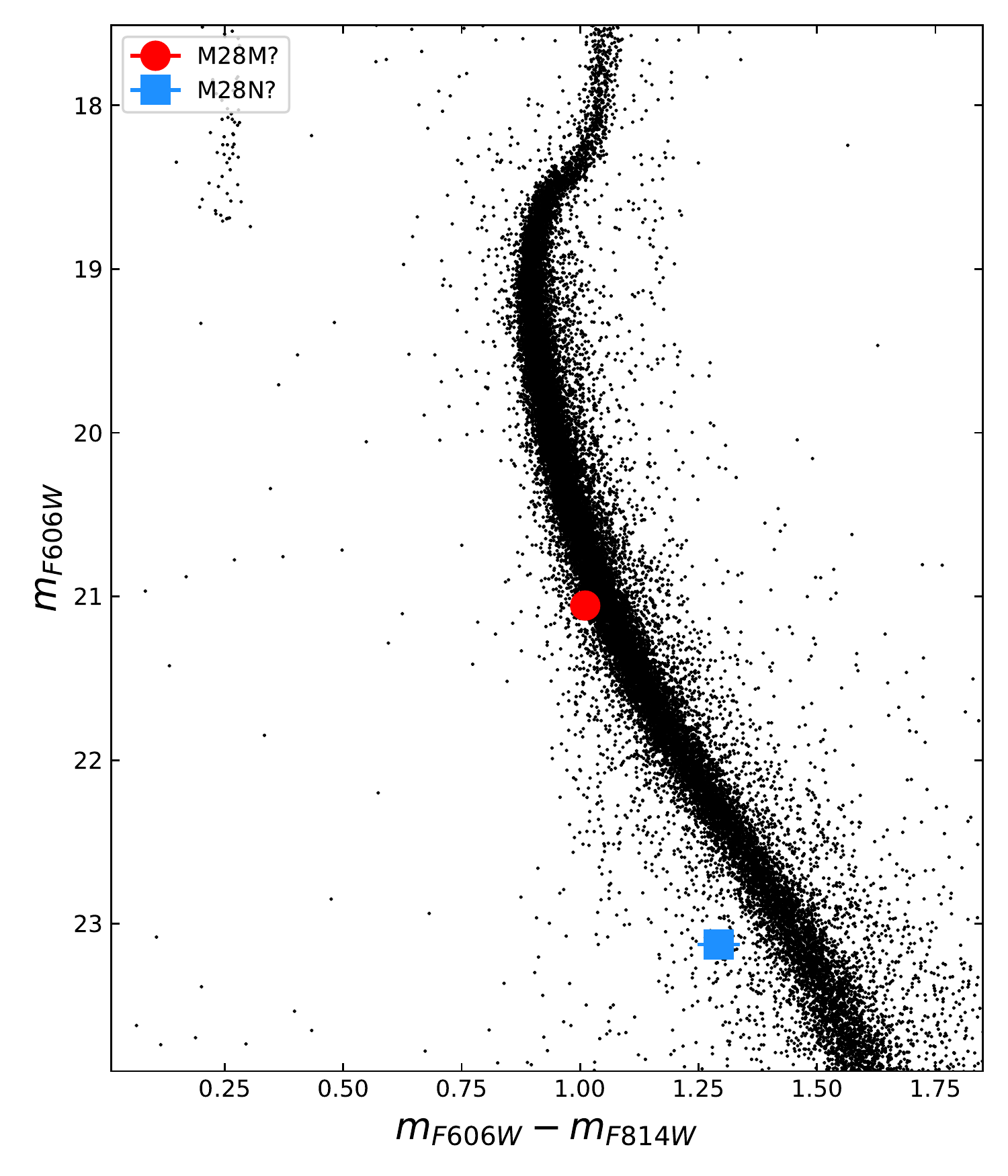}

\caption{{\it Left-hand panel:} CMD of M28 obtained in the F390W$-$F814W filter combination. Magnitudes have been corrected for the effects of differential reddening following the prescriptions by \citet{cadelano20a}. The positions of the candidate counterparts to M28M and M28N are highlighted with a red point and a blue square, respectively. {\it Right-hand panel:} same as in the left-hand panel, but in the F606W$-$F814W filter combination.}
\label{fig:cmd}
\end{figure}

In the color-magnitude diagrams (CMDs), the candidate companion to M28M is located along the cluster main sequence, $\sim2.9$ mags below the turn-off (see Figure~\ref{fig:cmd}). Interestingly, it shows a hint of red excess in the F390W(U-band)$-$F814W(I-band) filter combination, while a small blue excess is detected in the F606W(V-band)$-$F814W combination. The observations in the $H_\alpha$ narrow filter (i.e. F658N) show no sign of $H_\alpha$ excess.

 On the other hand, the candidate companion to M28N is located at $\sim5.4$ mag below the turn-off and shows a blue excess in all the available filter combinations. Such an excess is commonly observed in exotic binaries like cataclysmic variables and MSPs \citep[e.g.][]{riverasandoval18,zhao19,zhao20}, and it is usually related to the presence of an accretion disk and/or heating of the secondary star by the primary. Also in this case, no evidence of $H_\alpha$ excess is detected.
 
 Despite the good positional agreement and the peculiar CMD positions, both the objects show no sign of photometric variability in any of the available filters. However, companions to black widows identified both in globular clusters and in the Galactic field are often characterized by strong magnitude variations over their short orbital periods. The phenomenon is commonly due to the heating of the companion side exposed to the pulsar wind \citep[e.g.][]{pallanca14,cadelano15,kaplan18,zharikov19}. The absence of any magnitude modulation in the proposed candidate companions might suggest that either the two objects are not associated with the pulsars or the heating phenomenon is significantly less efficient than observed in typical black widow systems (possibly also due to geometrical effects).
 

Given all this, we cannot safely conclude that the two proposed counterparts are associated with the MSPs. In fact, the offset positions with respect to the cluster main-sequence might also indicate that these two objects are field interlopers rather than cluster members. Such a possibility could be confirmed in the future through their proper motion analysis. Moreover, the only two black widow optical counterparts identified in a globular cluster so far (i.e. PSR~J1518$+$0204C and PSR~J1953$+$1846A; \citealt{pallanca14,cadelano15}) are located 6 to 9 magnitudes below their cluster turn-off (depending on the orbital phase) in the F606W filter. Therefore, both the objects are fainter than the two optical counterparts here presented. This strengthens the likelihood that these two objects are physically unassociated with the pulsars but are unrelated stars belonging either to the M28 population or to the foreground.


\subsection{Searching for x-ray counterparts}

While we did not perform an independent analysis of the multiple {\em Chandra} observations of M28, we did query the online {\em Chandra} Source Catalog (CSC) v2.0 to see if any point sources matched the positions of the two pulsars.  Since the pulsars are only separated by 3.7\arcsec, the two closest sources to each pulsar turn out to be the same two CSC sources, 2CXO~J182433.0$-$245210, which was source 29 in \citet{Becker2003}, and 2CXO~J182433.2$-$245209, although neither x-ray source is within 1.7\arcsec (more than twice the radius of the formal x-ray positional errors) of the timing position of either pulsar.  The limiting sensitivity of the combined observations at the pulsar positions is $\sim$1.2$\times$10$^{-15}$\,erg\,s$^{-1}$cm$^{-2}$\, for M28M and $\sim$1.0$\times$10$^{-15}$\,erg\,s$^{-1}$cm$^{-2}$\ for M28N, in the ACIS 0.5$-$7\,keV band, corresponding to luminosity limits of $\lesssim$4.0$\times$10$^{30}$\,erg\,s$^{-1}$ and $\lesssim$3.4$\times$10$^{30}$\,erg\,s$^{-1}$ for M28M and M28N, respectively, at the 5.37\,kpc distance of M28 \citep{bv2021}.  These limits are similar to the measured ACIS-band luminosities of the many detected millisecond pulsars, including several black widows, in the globular cluster 47~Tucanae \citep[e.g.][]{bogdanov2006,Bhattacharya2017}, and are therefore not exceptionally constraining.

\section{Discussion}
\label{sec:discussion}
There are many interesting implications from the timing of both M28M and M28N. In this section, we compare the exotic binary population in M28 to that in other globular clusters, we examine the implications of orbital period derivative measurements (or limits), and we compare the timing-based proper motion measurements with the Gaia-based proper motion of M28 as a whole.

\subsection{The ``Spider'' population}
\label{section:exotic_bin} 

A peculiarity of M28 is its significantly larger fraction of spiders (see Section \ref{sec:intro}) than 47~Tucanae (hereafter 47Tuc) and Terzan~5 (hereafter Ter5), the two globular clusters with the richest content in pulsars to date. In particular, introducing the {\it ratio $\mathcal{R}$} between the population of spiders and the total known pulsar population in a globular cluster, we have $\mathcal{R}_{\rm M28}=50\%$ (14 known pulsars, among which are five black widows and two redbacks \citep{beg06}), $\mathcal{R}_{\rm 47Tuc}=33\%$ (27 known pulsars with six black widows and three redbacks), and $\mathcal{R}_{\rm Ter5}=13\%$ (two black widows and three redbacks out of a total of 39 known pulsars). Summing over the data for all the other 33 globular clusters containing known pulsars, the ratio is $\mathcal{R}_{\rm 33gc}=14\%$, close to the value seen in Ter5\footnote{All the numbers are taken from \url{https://www.naic.edu/~pfreire/GCpsr.html}}. However, we focus only on the three aforementioned most populated clusters, as we believe that the sample size for them is large enough that the fraction of spiders can be seen as a trend, at least to first approximation.  

Using our knowledge of the numbers of spiders in M28, 47Tuc and Ter5, we can make a guess as to how many of these types of systems might be potentially observable in the three clusters. \citet{Bagchi2011} utilized Monte Carlo simulations to predict the number of observable pulsars using six different models. These models predicted M28 to have between 42 and 952 observable pulsars, with an average between the models of 320 $\pm$ 60. From the same analysis, Ter5 should have a model averaged pulsar content of 380 $\pm$ 40 pulsars and 47Tuc of 160 $\pm$ 20 pulsars. It must be noted that there is a large discrepancy in the absolute number of predicted pulsars across the range of these models, but the relative ratios of predicted pulsars in the 3 globular clusters are roughly compatible between models, with Ter5 predicted to have a total pulsar population 15\%-40\% larger than M28 and 120\%-150\% larger than 47Tuc. Combining these percentage intervals with the values of $\mathcal{R}$, we find that M28 should be by far the leading globular cluster in terms of absolute number of spiders, containing about three times the number of spiders than either of the other two globular clusters, the spider content of which should be similar to each other. In absolute terms, according to the entire range of models explored by \citet{Bagchi2011}, M28 might host from $\sim 20$ observable spiders up to $\sim 400$, while assuming a conservative log-normal distribution in the pulsar luminosity, the simulations of \citet{Bagchi2011} indicate a number of potentially observable spiders in M28 ranging from $\sim 20$ to $\sim 80.$

These results naturally raise the question about the reasons for the overabundance of spiders in M28 with respect to other globular clusters, and in particular 47Tuc and Ter5.  It is well known that the formation of spiders in a globular cluster can occur via two channels: ordinary binary evolution (as in the Galactic field), or through the occurrence of 3- or 4-body dynamical interactions, which can support the creation of new binaries with suitable parameters to later become spiders.  The former pathway certainly depends on the total mass $M_{\rm GC}$ of the globular cluster, while the latter process is in turn favored by a crowded and massive stellar environment, suggesting an exploration of both the role of the rate $\Gamma_c$ of stellar interactions in the core (which depends on the central density, $\rho_c$ and the core radius of the globular cluster, $r_c$, $\Gamma_c=\rho_c^{3/2}r_c^{2}$, see e.g.~\citealt{Verbunt2014}), and of the mass density within the half-mass radius $\rho_{hm}$.  Using the data provided in the website by Holger and Baumgardt\footnote{Structural Parameters of Globular Clusters: \url{https://people.smp.uq.edu.au/HolgerBaumgardt/globular/parameter.html}}, as far as the values of the total mass, Ter5, 47Tuc, and M28 (sorted in this way) appear in the top 23\% of the listed globular clusters. They are also in the top 22\% in terms of interaction rate in the core (with Ter5 leading over M28 and 47Tuc). As for $\rho_{hm},$ M28 (sixth position) and Ter5 (ninth position) fell in the top 10\% of the globular clusters, with 47Tuc in the top 27\%. The high ranking in all the mentioned lists is in agreement with the observed occurrence of a high number of millisecond pulsars in the three clusters. However, M28 prevails among the three clusters only in terms of mass density within the half-mass radius, with a limited edge on Ter5, i.e.  $\log\rho_{hm,\rm M28}=3.49,$ and $\log\rho_{hm,\rm Ter5}=3.34.$ On the other hand, despite a similar predicted number of spiders (see above), Ter5 and 47Tuc show significantly different values of the latter parameter ($\log\rho_{hm,\rm 47Tuc}=2.65$), as well as of $\Gamma_c$ (three times larger in Ter5 than in 47Tuc), having only similar values for the total mass. 

In summary, besides the suggested impact of $\rho_{hm}$, which could be explored with detailed simulations of its effects on the production of the spiders, some other still unassessed and/or as yet unmeasured parameters are at play in determining the population of spiders in a cluster, e.g.~one can think about the fraction of binaries, or the past dynamical history of the cluster. An interesting indication could arise from the distinct rarity of eclipses in the sample of the M28 spiders. In fact, of the five black widow systems, only one shows obvious eclipses, M28G \citep{beg06}. The timing analyses of these systems also do not display strong orbital variability, which is very commonly detected in similar systems. In principle, these peculiarities could  be due to a geometrical effect, i.e.~we are looking onto the systems nearly face on, or at an inclination of near $0\degrees$. However, this is statistically unlikely to occur for many binary systems at once. An alternate option is that the relative absence of eclipses and the orbital stability are indications that the companion star is stable and there is little stellar wind being ablated. This often implies that the systems have had time to stabilize and hence are older than the typical population of spiders in globular clusters. If this hypothesis is supported by additional observations and further discoveries in M28, it would suggest that the high number of spiders in this globular cluster is sustained by an accumulation mechanism, where the spiders were produced far in the past yet have managed to survive intact in the cluster environment. 

\begin{center}
\begin{deluxetable*}{lccc}
\tablewidth{0pt}
\tablecaption{Comparative Projected Pulsar Populations in M28, 47Tuc, and Ter5}
\startdata
\tablehead{\colhead{Cluster Properties} & \colhead{M28}  & \colhead{47 Tucanae} & \colhead{Terzan 5}}
Average Potential Observable MSPs \dotfill & 320 $\pm$ 60 & 160 $\pm$ 20 & 380 $\pm$ 40\\
Ratio of "Spider" Systems To Known MSPs ($\mathcal{R}$) \dotfill & 0.50   &  0.33  & 0.13\\
Mass Density Within Half-mass Radius ($\log\rho_{hm}$) \dotfill & 3.49 & 2.65 & 3.34\\
\enddata\
\tablecomments{These are the relevant values taken from section 4.1
\label{table:poplations}}
\end{deluxetable*}
\end{center}

\subsection{Corrections for Pulse Period Derivative}
\label{sec:accels}
Pulsars in Globular clusters are often found to have seemingly nonphysical values of $\dot{P}$ compared to pulsars in the Galaxy. This contamination is a result of primarily the acceleration of the pulsar in the gravitational field of the cluster and leaves no vestige of the intrinsic period derivative in the observed period derivative \citep{Phinney1992}. In order to solve for the intrinsic period derivative \(\dot{P}_{\rm int}\), we can \citep[e.g.][]{Freire2017} use the equation
\begin{equation}
    \frac{\dot{P}_{\rm obs}}{P} = \frac{\dot{P}_{\rm int}}{P} + \frac{\mu^2 d}{c} + \frac{a_{l,{\rm GC}}}{c} +\frac{{a}_{l,{\rm gal}}}{c},
    \label{eqn:accel}
\end{equation}
where \(\dot{P}_{\rm obs}\) is the observed spin-period derivative. The second term on the right represents the Shklovskii effect \citep{1970shk} due to the motion of the pulsar in the plane of the sky; $\mu$ is the proper motion of the cluster, $d$ is the distance to the cluster, and $c$ is the speed of light. The third term, $a_{l,{\rm GC}}$ represents the line-of-sight acceleration of the pulsar in the globular cluster. The fourth term represents the difference in acceleration, $a_{l,{\rm gal}}$, between the solar system and the cluster in the field of the galaxy.

The Shklovskii effect may be calculated as $3.12 \times 10^{-10}$\,m\,s$^{-2}$ using the total proper motion of the cluster (8.9 mas/year) and its 5.37\,kpc distance \citep{vb2021,bv2021}). The galactic acceleration may be computed as $3.08 \times 10^{-10}$\,m\,s$^{-2}$ using the galactic coordinates of M28 and its distance \citep[e.g.][]{Freire2017}. The line-of-sight acceleration from the cluster may then be solved for using
\begin{equation}
    a_{l,{\rm GC}} = \frac{\dot{P}_{b,{\rm obs}}}{P_{b}} c - \mu^2 d - a_{l,{\rm gal}}
\end{equation}
\noindent where $P_b$ and $\dot{P}_{b,{\rm obs}}$ are the orbital period and its observed period derivative.  For M28N, which has a more significant measurement of $\dot{P}_{b,{\rm obs}}$, the cluster acceleration is $1.02 \times 10^{-8}$\,m\,s$^{-2}$.  With that value, we can then solve for $\dot{P}_{\rm int}$ with Eqn.~\ref{eqn:accel}, giving $\sim3.8 \times 10^{-20}$ for M28N, and allowing estimates of the spin-down based physical parameters in Table~\ref{table:timing}. 

With a less precise measurement of $\dot{P}_{b,{\rm obs}}$ for M28M, we can assume a 95\% lower limit of the true $\dot{P}_{b,{\rm obs}}$ of $4.4 \times 10^{-13}$ given the estimated error on $\dot{P}_{b,{\rm obs}}$ and then proceed in a similar manner for M28N to get an upper limit of $\dot{P_{\rm int}} < 2.2 \times 10^{-20}$ with corresponding upper limits (magnetic field and $\dot{E}$) or lower limits (characteristic age) on the resulting physical spin-down parameters in Table~\ref{table:timing} (denoted by \dag).

\subsection{Proper Motions}
\label{section:pm}
 The long timing baseline provided by the GBT data allows low-precision estimates of the individual pulsar proper motions as well as the orbital period derivatives mentioned in the previous section. The Gaia mission has measured the proper motion of M28 as $\mu_{\alpha} = -0.28(3)$\,mas\,yr$^{-1}$ and $\mu_{\delta} = -8.92(3)$\,mas\,yr$^{-1}$, for a total proper motion of 8.92(3)\,mas\,yr$^{-1}$ \citep{vb2021}).  The timing-based proper motions (or limits on measurements) are consistent with the Gaia values within one sigma, supporting the association of both binary systems with M28 and their being bound to the cluster.

\section{Summary and ramifications}
\label{sec:conclusion}

We used sensitive new TRAPUM and MeerTime data to detect two new black widow pulsars in the globular cluster M28, and exploited archival GBT observations of the cluster to determine long-term coherent timing solutions for both binary pulsars.  Their orbital parameters reflect the typical values seen in black widow systems in other globular clusters. However, their discovery supports the growing evidence that the structural parameters and the internal dynamics of M28 make it a very effective factory of exotic pulsar binaries, here defined as the joint class of black widow and redback systems.  Half of the known pulsars in M28 belong to these categories, and projecting this high fraction on preceding pulsar population studies, M28 could harbor several tens or even some hundreds of these binaries. Some of them could be unveiled during the ongoing search of the outer beams of the TRAPUM observations.
We also searched for the companions (or overall systems) of M28M and M28N in the optical and X-ray bands, but were unable to conclusively identify either system in either band. 

Besides confirming the sensitivity and the potential of the MeerKAT telescope to find  new pulsars in globular clusters (with these discoveries the total count for MeerKAT reaches 36 new pulsars\footnote{\url{http://trapum.org/discoveries.html}}), the case of M28M and M28N also highlights the value of the long-term monitoring and archiving of globular cluster pulsar data. As in the case of Terzan5 \citep{Ridolfi2021}, the use of archival GBT data allowed us to establish for M28M and M28N a $\sim$ 16-year-long timing solution soon after the discovery of these pulsars. This is especially important for the data from large single-dish telescopes, like the GBT, which have wider beams and very likely include the positions of newly discovered pulsars.

In this context, it is worth noting that M28M and M28N were not discovered in the GBT data because they are, on most occasions, too faint to be detected by blind search algorithms. However, after their orbits had been characterized in the MeerKAT observations, the number of trials necessary for a detection decreased dramatically. That implied that the pulsed signals from these objects could be recovered in the archival GBT data with a signal-to-noise high enough for determining a suitable set of pulse times of arrival and then deriving a phase-connected timing solution with highly precise timing parameters. For these reasons, it is clear that the current archival data of many globular clusters will continue to be of great use for fast characterization of many MeerKAT discoveries. The same also applies, of course, for the archival data from HST and from high-energy telescopes, which are also important to immediately investigate the new pulsars at those wavelengths.

\begin{acknowledgments}
The MeerKAT telescope is operated by the South African Radio Astronomy Observatory, which is a facility of the National Research Foundation, an agency of the Department of Science and Innovation. SARAO acknowledges the ongoing advice and calibration of GPS systems by the National Metrology Institute of South Africa (NMISA) and the time space reference systems department of the Paris Observatory. MeerTime data is housed on the OzSTAR supercomputer at Swinburne University of Technology. The OzSTAR program receives funding in part from the Astronomy National Collaborative Research Infrastructure Strategy (NCRIS) allocation provided by the Australian Government. The Green Bank Observatory is a facility of the National Science Foundation operated under
cooperative agreement by Associated Universities, Inc.

SMR is a CIFAR Fellow and is supported by the NSF Physics Frontiers Center awards 1430284 and 2020265. AR, AP, and MB acknowledge the support from the Ministero degli Affari Esteri e
della Cooperazione Internazionale - Direzione Generale per la
Promozione del Sistema Paese - Progetto di Grande Rilevanza ZA18GR02. Part of this work has been funded using resources from the research grant “iPeska” (P.I. Andrea Possenti) funded under the INAF national call Prin-SKA/CTA approved with the Presidential Decree 70/2016. Pulsar research at UBC is supported by an NSERC Discover Grant and by the Canadian Institute for Advanced Research. AR acknowledges continuing valuable support from the Max-Planck Society. J.W.T.H. acknowledges funding from an NWO Vici grant ("AstroFlash"; VI.C.192.045). M.D. acknowledges support from the National Science Foundation (NSF) Physics Frontier Center award 1430284, and from several NASA grants at the Naval Research Laboratory. TRAPUM observations usedthe FBFUSE and APSUSE computing clusters for data acquisition,storage and analysis. These clusters were funded and installed bythe Max-Planck-Institut für Radioastronomie and the Max-Planck-Gesellschaft. L.V. acknowledges financial support from the Dean’s Doctoral Scholar Award from the University of Manchester.

\end{acknowledgments}
\facilities{MeerKAT, GBT}
\software{PRESTO \citep{ASCLPRESTO},
          PSRCHIVE \citep{ASCLPSRCHIVE},
          Astropy \citep{ASCLAstropy}}

\begin{center}
\begin{deluxetable*}{lccc}
\tablewidth{0pt}
\tablecaption{New MeerKAT Discovered M28 Pulsars}
\startdata
\tablehead{\colhead{Parameter} & \colhead{Value(Error)}  & \colhead{Value(Error)}}
Pulsar Name \dotfill & J1824$-$2452M   & J1824$-$2452N\\
Right Ascension (RA, J2000) \dotfill & $18^{\rm h}\;24^{\rm m}\;33\fs1835(5)$   & $18^{\rm h}\;24^{\rm m}\;33\fs1418(12)$\\
Declination     (DEC, J2000) \dotfill & $-24\degrees\;52\amin\;08\farcs2(23)$   & $-24\degrees\;52\amin\;11\farcs89(3)$\\
Proper Motion Right Ascension (mas/yr) \dotfill & $0.0(1.4)$   & $0.5(0.3)$\\
Proper Motion Declination     (mas/yr) \dotfill & $ -23(28)$   & $ -15(6)$\\
Pulsar Period (ms) \dotfill & 4.7842842982785(7)   & 3.35287243686099(10)\\
Pulsar Frequency (Hz) \dotfill & 209.01767906222(3)   & 298.251728579395(9)\\
Frequency Derivative (Hz\,s$^{-1}$)  \dotfill & $-5.36572(12) \times 10^{-15}$    & $-1.416227(3) \times 10^{-14}$\\
Frequency Second Derivative (Hz\,s$^{-2}$)   & $-1.4(4) \times 10^{-26}$    & $-9.4(1.2) \times 10^{-27}$\\
Reference Epoch (MJD) \dotfill & 56450   & 56450\\
Dispersion Measure (pc cm$^{-3}$) \dotfill & 119.352(11)   & 119.321(3)\\
Orbital Period (days) \dotfill & 0.2425192190(14)   & 0.19849331510(11)\\
Orbital Period Derivative  \dotfill & $3.1(1.6) \times 10^{-12}$   & $6.2(1.4) \times 10^{-13}$\\
Projected Semi-Major Axis (lt-s)  \dotfill & 0.032463(8)   & 0.0497303(16)\\
Time of Ascending Node  \dotfill & 56451.272704(15)   & 56451.2896713(15)\\
\cutinhead{Properties of Fit}
Span of Timing Data (MJD) \dotfill & 53629$-$59274    & 53629$-$59274\\
Number of TOAs  \dotfill & 161   & 230\\
RMS TOA Residual ($\mu$s) \dotfill & 71.11   & 16.45\\
\cutinhead{Derived Parameters}
Mass Function (\msun) \dotfill & $6.245(4) \times 10^{-7}$   & $3.3516(3) \times 10^{-6}$\\
Min Companion Mass (\msun) \dotfill & $\geq$\,0.011   & $\geq$\,0.019\\
Pulsar Magnetic Field  (G) $\dagger$   \dotfill   &  $<4 \times 10^{8}$ & $3.6 \times 10^{8}$ \\
Pulsar Age  (years) $\dagger$ \dotfill   & $>3 \times 10^{9}$ & $1.4 \times 10^{9}$ \\
Pulsar $\dot{E}$  (erg\,s$^{-1}$) $\dagger$ \dotfill    & $<8 \times 10^{33}$ & $4.0 \times 10^{34}$ \\
Flux Density at 2\,GHz ($\mu$Jy) \dotfill & 6(3)  & 10(5)\\
Angular Offset of Cluster Center (mas)  \dotfill & 5.092   & 3.464\\
Total Proper Motion (mas/yr) $\ddag$  \dotfill & 10(22)   & 14(7)\\
\enddata

\tablecomments{Numbers in parentheses represent 1-$\sigma$
uncertainties in the last digit.  The timing solution was
determined using {\tt TEMPO} with the DE440 \citep{park2021} Solar System Ephemeris and
the ELL1 binary model. The time system used is Barycentric Dynamical
Time (TDB), referenced to TT via BIPM.  The minimum companion mass was calculated assuming a pulsar mass of 1.4\,\msun. $\dagger$ Here we assumed a 95\% lower limit to $\dot{P}_{b}$ which allowed us to solve for an upper limit on the intrinsic $\dot{P}$ and therefore a lower limit on the age and upper limits on the magnetic field and $\dot{E}$ (see \S\ref{sec:accels}). $\ddag$ The relatively poorly constrained proper motions are nevertheless consistent with overall cluster proper motion measured by Gaia (see \S\ref{section:pm}).
\label{table:timing}}
\end{deluxetable*}
\end{center}

\bibliography{Twonewblackwidowpulsars}{}
\bibliographystyle{aasjournal}



\end{document}